\documentclass[sigconf, nonacm]{acmart}

\usepackage{fancyvrb}
\usepackage{tabularx}
\usepackage{float}
\usepackage{balance}
\usepackage[inline]{enumitem}

\setcopyright{none}
\usepackage{siunitx}
\AtBeginDocument{%
  }

\setlist[itemize]{leftmargin=*, itemsep=2pt}

\begin{document}

%\title{Generative Retrieval Overcomes the Limitations of Dense Retrieval}
%\title{Beyond Vector Space: How Generative Retrieval Solves High-Rank Relevance}
\title{Generative Retrieval Overcomes Limitations of Dense Retrieval but Struggles with Identifier Ambiguity}


% --- Anonymous authors ---
%\author{Anonymous Author(s)}
%\affiliation{
%  \institution{Paper under double-blind review}
%  \city{Anonymous City}
%  \country{Anonymous Country}
%}
%\email{anonymous@acm.org}

\author{Adrian Bracher}
\email{adrian.bracher@wu.ac.at}
\orcid{0009-0007-0956-0388}
\affiliation{%
  \institution{Vienna University of Economics and Business}
   \city{Vienna}
   \country{Austria}
 }
 \author{Svitlana Vakulenko}
 \email{svitlana.vakulenko@wu.ac.at}
 \orcid{0000-0002-5278-8886}
 \affiliation{%
   \institution{Vienna University of Economics and Business}
   \city{Vienna}
   \country{Austria}
 }

\begin{abstract}
%Dense retrieval models using fixed-dimensional embeddings geometrically limit the data they can accurately represent~\cite{limit, reimers2021curse}. While sparse models like BM25 escape these specific constraints, they fail to capture semantics beyond exact term matching, highlighting a fundamental representational trade-off.

While dense retrieval models, which embed queries and documents into a shared low-dimensional space, have gained widespread popularity, they were shown to exhibit important theoretical limitations and considerably lag behind traditional sparse retrieval models in certain settings.
Generative retrieval has emerged as an alternative approach to dense retrieval by using a language model to predict query-document relevance directly.
In this paper, we demonstrate strengths and weaknesses of generative retrieval approaches using a simple synthetic dataset, called LIMIT, that was previously introduced to empirically demonstrate the theoretical limitations of embedding-based retrieval but was not used to evaluate generative retrieval.
We close this research gap and show that generative retrieval achieves the best performance on this dataset without any additional training required (0.92 and 0.99 R@2 for SEAL and MINDER, respectively), compared to dense approaches (< 0.03 Recall@2) and BM25 (0.86 R@2).
However, we then proceed to extend the original LIMIT dataset by adding simple hard negative samples and observe the performance degrading for all the models including the generative retrieval models (0.51 R@2) as well as BM25 (0.21 R@2). Error analysis identifies a failure in the decoding mechanism, caused by the inability to produce identifiers that are unique to relevant documents. Future generative retrieval must address these issues, either by designing identifiers that are more suitable to the decoding process or by adapting decoding and scoring algorithms to preserve relevance signals.

\end{abstract}





\maketitle

\section{Introduction} \label{sec:introduction}

% representational capacity, historical shift from sparse to dense retrieval due to improved representation of semantics
%The evolution of Information Retrieval (IR) has been driven by the pursuit to efficiently and accurately encode mappings between queries and relevant documents, i.e., enhancing representational capability. Traditional sparse methods, such as BM25~\cite{bm25}, rely on exact term matching to represent relevance. While efficient, these approaches suffer from the \textit{vocabulary mismatch problem}, effectively failing to retrieve relevant documents that do not share the exact terms occurring in the query. Dense retrieval (DR) addresses this limitation by embedding queries and documents into a shared vector space for similarity-based retrieval. Thereby, DR methods can represent semantics independent of the terminology, a generational paradigm shift in modern IR.

% !!! describing DR and its limitations
Dense Retrieval (DR) models, characterized by their ability to capture latent semantics beyond surface-level term matching, fueled the recent generational paradigm shift in modern Information Retrieval (IR)~\cite{dpr, contriever, promptriever}.
However, recent studies indicate that compressing information into fixed-dimensional vector spaces — a core tenet of DR — imposes fundamental geometric constraints on the resulting data representations.
Specifically, as the volume of information and the complexity of queries grows, the limited representational capacity of these dense vector embeddings leads to a ``vector bottleneck''~\cite{limit}.
This occurs when the model's ability to distinguish between relevant and irrelevant documents by partitioning the embedding space into distinct, relevant subsets is exhausted, leading to representation collapse and a clear degradation in retrieval performance~\cite{reimers2021curse,fayyaz2025collapse}.

\begin{figure}[h]
  \centering 
  \includegraphics[width=\linewidth]{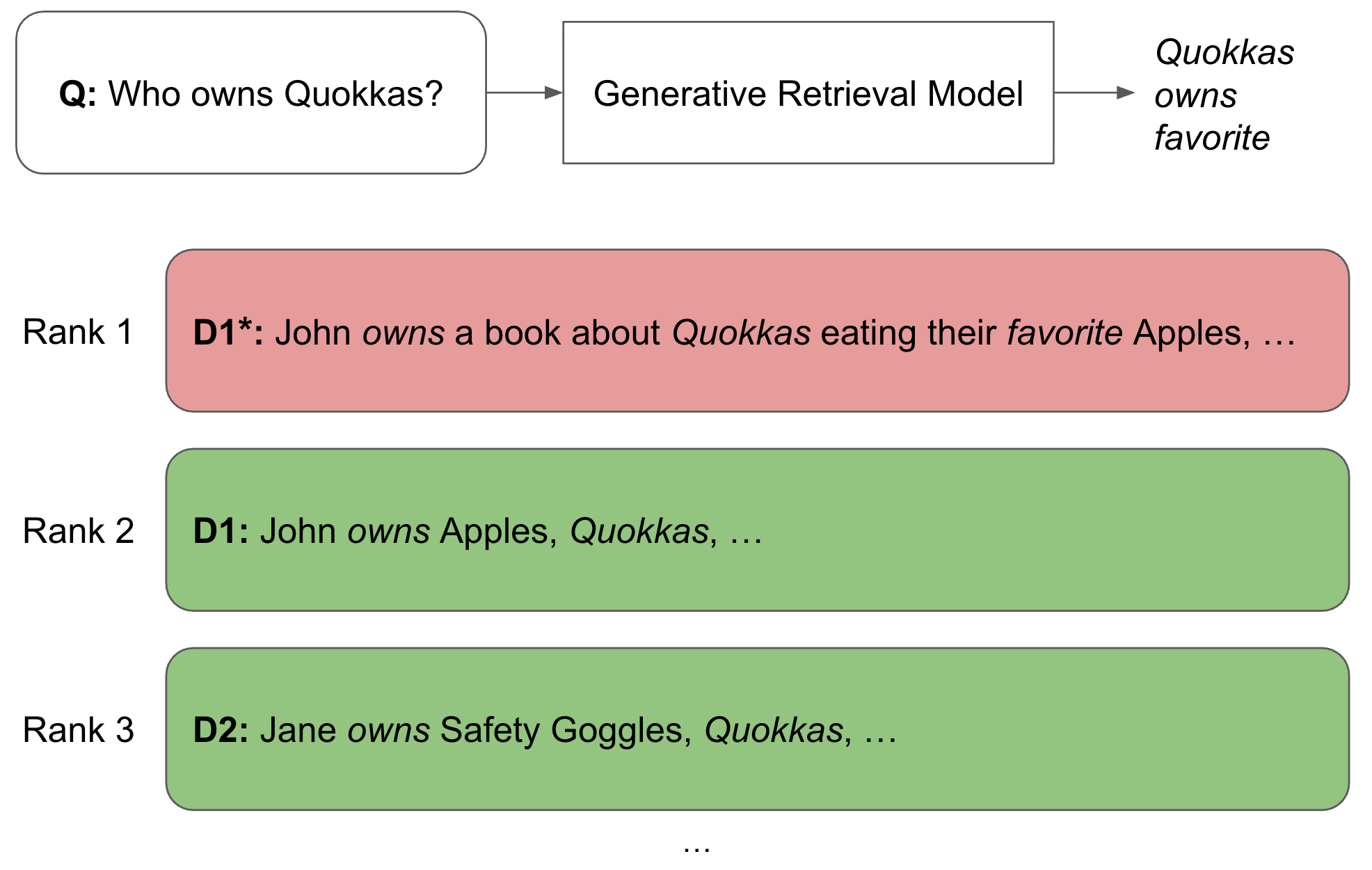}
  \caption{An example from LIMIT-H demonstrating ambiguity of the generated document identifiers (\textit{italic}) that prevent the GR model from distinguishing between relevant (green) and irrelevant (red) documents.} 
\label{fig:ambiguous_docid_retrieval}
\end{figure}

% % GR potential unifier that encodes relevance in parameters (theoretically solving 1) while retaining the deep learning capacity for (2)
% However, in current GR successful retrieval is contingent on docids that are suitable to the sequential decoding process and capture semantics of documents.

% !!! describing GR and its potential
Generative Retrieval (GR) has emerged as an alternative to DR that utilizes an autoregressive language model (LM) to generate relevant document identifiers (docids) given only a query as input~\cite{dsi, seal, nci}.
Unlike the vector-space computations in DR, GR aims to encode complex relevance relationships directly within the LM's parameters.
\citet{does_generative_retrieval_overcome} suggested that this architecture provides GR with superior representational power in comparison with DR and can potentially enable it to bypass DR's limitations.

Intrigued by the theoretical advantages of GR, we conduct an empirical evaluation using the same benchmark that exposed the fundamental limitations of state-of-the-art DR models.
The LIMIT dataset, introduced by \citet{limit}, is a synthetic benchmark designed to be minimalist, yet it proved surprisingly challenging in practice, allowing the authors to effectively quantify the ``vector bottleneck.''
Our study investigates whether the parameter-based memory of GR can naturally navigate around this geometric constraint by leveraging the high-dimensional internal state of an autoregressive model.

% !!! argue evaluation setup: only GR zero-shot approaches to avoid overfitting LIMIT which is trivial
Given that LIMIT is a synthetic benchmark specifically designed to expose structural representational limits, fine-tuning on it would allow a model to simply memorize and overfit to the specific patterns of the dataset rather than demonstrating true representational capacity.
Consequently, to ensure a rigorous and fair comparison with the original DR results reported by \citet{limit}, we restrict our evaluation to already pre-trained GR models — specifically SEAL and MINDER — that were fine-tuned on Wikipedia.
SEAL and MINDER are simple but high-performing GR models, that are often used as standard baselines for evaluations and are extended by more recent GR models~\cite{ltrgr,mekonnen2025ddro}.
Crucially, SEAL and MINDER were also shown to generalize well to documents unseen during training~\cite{li2024mdgr,mdgr}.
Therefore, we select them for our experiments as the representatives of the current state-of-the-art GR approaches.
By evaluating these models in a strictly zero-shot setting, we test whether the parametric memory of a generative model develops a mechanism for correctly identifying the query-document relationship that is robust and generalizable to out of distribution samples.

% !!! report results on original LIMIT: solved!
Our experiments show that despite the absence of any corpus-specific training, zero-shot GR models achieve near-perfect performance on the LIMIT benchmark.
While state-of-the-art dense retrievers (DR) exhibit representation collapse, failing to resolve the combinatorial constraints of the dataset, GR architectures successfully navigate these boundaries.
This stark contrast implies that the transition from vector-space matching (geometric similarity) to parametric memory retrieval (generative identifiers) may be essential for handling increasingly complex, multi-constraint information needs.

% By focusing on the LIMIT dataset, you are essentially testing whether moving relevance from the geometric space (vectors) to the model's internal weights (parameters) actually solves the scaling problem or simply moves the bottleneck elsewhere.

% Our results reveal a nuanced picture: while GR shows higher absolute capacity on the LIMIT benchmark, it exhibits a distinct set of failure modes related to identifier hallucination and sensitivity to prompt phrasing, suggesting that the 'parameter bottleneck' may simply be the next frontier to overcome.

% While recent work has argued ..
% limitations of that paper showing the gap and our contribution to filling this identified gap
% In this paper, we ...
% !!! high-level description of LIMIT

% An example from LIMIT is shown in Figure~\ref{fig:ambiguous_docid_retrieval}.
% This dataset constructed to exhaust all pairwise combinations of a document subset.
% Each document contains a statement about a person liking 50 items.
% Queries correspond to selecting a single item.

% In LIMIT, there are always two relevant documents for each query, i.e., exactly two people like the same item.
% This combinatorial enumeration enforces geometric conditions that prohibit separation by fixed-dimensional spaces.
% On this task, state-of-the-art (SoA) single-vector DR models achieve less than 0.03 Recall@2, whereas BM25, which is not constrained by embedding dimensionality, achieves 0.86 R@2.

% LIMIT-H
While the original LIMIT dataset effectively benchmarks robustness against geometric constraints, it does not account for the \textbf{semantic ambiguity} that is more prevalent in real-world IR.
For instance, while BM25 outperforms DR on LIMIT by a large margin, it relies on surface-level term matching and will struggle to resolve complex semantic relationships, such as negation or homographs.

% LIMIT-H
To test IR models against structural and semantic limitations simultaneously, we introduce \textbf{LIMIT-H}.
In this variant, we augment the original dataset with ``hard negative'' samples (D1*) that lexically overlap with the positive samples (D1) but carry different meanings and are irrelevant to the original queries (see Figure~\ref{fig:ambiguous_docid_retrieval} for an illustrative example from LIMIT-H).
This minimum change requires the model to not only navigate the vector bottleneck but also demonstrate the representational capacity for fine-grained semantic discrimination between the relevant documents and the semantic distractors we introduced.

While GR outperformed all other IR models on the original LIMIT benchmark, our experiments on LIMIT-H show that both BM25 and GR fail as semantic ambiguity increases. 
These results provide empirical evidence that GR architectures are capable of overcoming the representational limitations observed in DR models.
However, while GR has the potential to capture semantics using a LM pre-trained on a general corpus, our findings suggest that the current approaches we tested are not sufficient to take on this challenge identifying an important research gap.

% this capability can only be realized with robust document identifiers.

%Specifically, when faced with negatives designed to exploit the ambiguity of the identifier scheme, we observe that R@2 degrades from the original 0.988 to 0.690 on LIMIT-H. Moreover, a higher density of negative documents reduces performance to 0.571, demonstrating that the model's vulnerability worsens as the degree of identifier ambiguity increases.

% contributions:
% empirical evidence that GR solves challenge 1 better than all other methods
% show that GR can only be as good as docid design allows
% empirical results

In summary, we make the following research contributions:
\begin{enumerate}
\item{\textbf{Empirical validation of GR parametric capacity}: We provide strong empirical evidence that GR architectures are structurally superior to DR in zero-shot settings. By leveraging high-dimensional parametric memory to generate discrete identifiers, GR successfully overcomes the vector bottleneck introduced by the LIMIT benchmark, achieving near-perfect retrieval.} 
\item{\textbf{Introduction of the LIMIT-H benchmark}: We introduce LIMIT-H, a challenging extension of the LIMIT dataset that incorporates semantic ambiguity designed to test a model's dual ability to handle structural complexity and fine-grained semantic discrimination.}
\item{\textbf{Revealing GR limitations}: Our analysis uncovers a hidden failure mode of popular GR models like SEAL and MINDER that generate non-unique identifiers: an inherent identifier ambiguity. While these models solve the geometric bottleneck, their reliance on substrings or pseudo-queries as identifiers makes them susceptible to ``semantic collisions'' when different documents share similar lexical views that result in incorrect relevance scoring for more fine-grained queries.}
\end{enumerate}

%  Our evaluation reveals that this simple benchmark poses an important challenge for all existing IR paradigms.
%In a zero-shot setting on the LIMIT benchmark, GR achieves 0.988 R@2.}

%\item{We demonstrate empirically that the semantic capability of GR is conditional on identifier uniqueness. Our results show that introducing just 46 negatives causes performance to degrade to 0.690 R@2, with a further drop to 0.571 when increasing ambiguity with 92 negatives, showing the need for more robust document identifier designs.}

\section{Related Work}
\label{sec:related_work} \label{sec:related_work}

% In this section we review other works that investigate limitations in information retrieval. We cover compression limitations and semantic limitations.
% \subsection{Compression Limitations}

Previous research exploring the theoretical limitations of DR and GR is relatively scarce.
% !! first paper to talk about DR limitations
To the best of our knowledge, \citet{reimers2021curse} were the first to formally define and discuss the theoretical limitations of DR, terming this phenomenon the ``curse of low-dimensionality.''
They identified a fundamental compression bottleneck in DR, arguing that fixed-dimensional embeddings impose a hard upper bound on the number of distinct documents a DR model can accurately represent.
Their empirical evaluation demonstrated that as index size increases, retrieval performance degrades even more drastically than the theoretical lower bound suggests.
They attributed this to the fact that neural embeddings do not efficiently utilize the full vector space but tend to cluster in narrow, high-density regions, further exacerbating the representational bottleneck.

% !! our main inspiration: LIMIT
Our main inspiration for this paper stems from the recent report by \citet{limit}, who demonstrated that retrieval performance is constrained not only by index size but also by task complexity w.r.t. the rank of the query-document relevance matrix.
They provided a formal proof for the low-rank bottleneck in DR.
As a primary example, they introduced the task where queries are specifically designed to represent all possible combinations of relevant document pairs.
To empirically validate this, they introduced the LIMIT benchmark, showing that despite its relatively small size, it still remains unsolvable for state-of-the-art DR models.
Similar to \citet{reimers2021curse}, they also observed that DR models fail to achieve a uniform distribution of embeddings and their effective dimensionality is much lower than the nominal vector dimension, which makes the vector bottleneck much narrower in practice and explains why DR models collapse even on relatively small but complex datasets like LIMIT.

The LIMIT benchmark also highlighted a surprising result: traditional sparse term-document models, such as BM25, significantly outperform state-of-the-art DR models on such complex tasks.
Since sparse models operate in a high-dimensional symbolic space, they are naturally more robust to the low-rank bottlenecks that plague dense embeddings.
In this paper, we extend this evaluation to GR models because it represents the next generation of sparse retrieval.
GR generates discrete, symbolic identifiers (like BM25 terms) but it does so using the deep contextual reasoning of a LM able to learn complex query-document relevance relationships.

% !! MdR
Similar to our findings, the concurrent work by \citet{does_generative_retrieval_overcome} argues for the fundamental advantages of GR over DR, citing its superior representational capacity.
However, their evaluation primarily targets fine-tuned models, which essentially tests the model's ability to memorize a specific document collection.
Their additional experiment shows a considerable drop in performance from a Hit@10 of 63.5 when fine-tuned to just 23.8 in a zero-shot setting.
In contrast, we investigate the generalization potential of GR.
Through a controlled experiment, we specifically focus on the models' inherent ability to generate suitable identifiers for previously unseen documents. Our results demonstrate that while GR successfully generalizes in a structurally complex scenario, it fails to distinguish between lexically similar documents due to identifier ambiguity.

\section{Preliminaries}
\label{sec:preliminaries}

In this section, we discuss the main concepts used in GR and provide more details specifically on SEAL and MINDER architectures, while arguing how their relevance ranking approach makes them uniquely suitable for our zero-shot experiments on the LIMIT benchmark.

% GR in detail
\subsection{Generative Retrieval}
Generative Retrieval (GR) departs from traditional ``matching'' IR paradigms by framing retrieval as a sequence-to-sequence task~\citep{dsi}.
Instead of calculating similarities in a vector space as in DR, a GR model is a LM with parameters $\theta$ trained to generate a relevant document identifier $n$ given a query $q$.
Each document identifier $n$ is a sequence of tokens (at least one) from the LM's vocabulary.
The probability of a document identifier $n$ is the product of conditional probabilities for each of its tokens $y_t$
\begin{equation}
P(n|q; \theta) = \prod_{t=1}^{|n|} P(y_t | y_{<t}, q; \theta)
\end{equation}
Many GR approaches also use another LM with a different set of parameters $\theta'$ to assign a docid $n$ to every document $d$ based on its content:
\begin{equation}
P(n|d; \theta') = \prod_{t=1}^{|n|} P(y_t | y_{<t}, d; \theta')
\end{equation}

% We stipulate that the effectiveness of a GR model, especially in a zero-shot setting, is largely determined by the nature of its docids.
The majority of GR models use synthetic docids, which are usually arbitrary numeric identifiers (DSI~\cite{dsi}, MDGR~\cite{mdgr}, PAG~\cite{pag}, RIPOR~\cite{ripor}).
This approach requires heavy supervision for the model to learn to assign semantics to these docids.

An alternative approach is to use \textbf{semantic} docids, such as questions, titles or substrings from the documents.
These docids either occur naturally, such as titles, URLs and tokens, or are produced by a supervised model, such as questions.

GR models are pre-trained on paragraph-documents but documents in the LIMIT benchmark contain single sentences.
To mitigate this distribution shift, we opt for an \textbf{unsupervised} approach in our experiments that was introduced in SEAL~\cite{seal}, which treats all ngrams of a document as its docids.
We also consider an extension to SEAL, called MINDER~\cite{minder}, that uses questions in addition to ngrams.

\subsubsection{SEAL~\cite{seal}}
\label{sec:seal}

The main approach behind SEAL is pairing a LM with an FM-index~\cite{fmindex}, which is an efficient data structure that stores full text of the original documents and enables efficient pattern matching over the whole collection.
It allows to determine the exact count and position of any ngram of length $m$ in $\mathcal{O}(m)$, irrespective of the collection size.

The FM-index in SEAL serves two main purposes:
\begin{enumerate}
    \item constrains LM decoding to ensure only the docids that actually exist in the documents as ngrams are generated;
    \item points to all the documents that contain those ngrams.
\end{enumerate}

Efficient decoding in SEAL is supported by a beam search algorithm, a common practice in GR~\cite{dsi}. However, SEAL modifies the standard pruning mechanism: rather than discarding candidates that fall outside the beam width, the model retains all partially decoded ngrams encountered during the search, along with their corresponding LM scores. Since these scores are computed during the expansion phase regardless of whether the candidate is selected, this retention incurs no additional computational cost. The advantage of this modification is finer-grained scoring by aggregating from a much larger set of ngrams than strictly those that survive the beam pruning.

For actual document ranking, SEAL uses \textit{LM+FM} score.
This score is a combination of the conditional probability, $P(n|q)$, produced by the LM for an ngram $n$ and the unconditional probability, $P(n)$, derived from the FM-index, which is similar to the document frequency (DF) of a term in the collection:
\begin{equation} \label{eq:ngram_score}
    w(n, q) = \textrm{max}\left(0, \textrm{log}\frac{P(n|q)(1-P(n))}{P(n)(1-P(n|q))}\right)
\end{equation}

For every document that contains at least one of the generated ngrams, we compute the document-level LM+FM score (Equation~\ref{eq:seal-document-level-score}) by aggregating the individual scores of the ngrams this document contains $K^{(d)}$:
\begin{equation} \label{eq:seal-document-level-score}
W(d, q) = \sum_{n \in K^{(d)}} w(n, q)^\alpha * \text{cover}(n, K^{(d)})
\end{equation}
where $cover$ is a function that handles duplicate ngrams and $\alpha$ is a hyperparameter for re-weighting the two terms.

% This set is constructed by (1) taking all generated ngrams that occur somewhere in the document, and then keeping only those ngrams that do not overlap with higher-scoring ngrams. 
% To prevent over-scoring of repetitive documents where the same ngrams are repeatedly matched each ngram is multiplied with a \textit{coverage weight}. $\alpha$ is a hyperparameter.

\subsubsection{MINDER~\cite{minder} }
This approach further builds upon SEAL and extends it by introducing additional docids (multi-view identifiers) to improve the original scoring.
Apart from ngrams, MINDER also assigns to each document two new types of docids:
\begin{enumerate}
    \item \textbf{Titles}: in our case those are skipped since they are absent in LIMIT;
    \item \textbf{Pseudo-queries}: synthetic questions generated by an auxiliary LM (e.g., Doc2Query~\cite{doc2query--}) given the document text.
\end{enumerate}

These new identifiers are then added to the FM-index preceded with a special token, \texttt{<TS>} and \texttt{<QS>}, to indicate the start of titles and pseudo-queries respectively.
During inference, the GR model is prompted with each of these special tokens separately to generate different types of docids (views).

To determine the final document rank, MINDER aggregates the evidence from all generated views, while each view is scored separately using the same LM+FM scores computations as in SEAL.
We consider MINDER in our experiments since it was designed to improve representational power of SEAL, particularly for the scenarios where a simple substring may be ambiguous.

%We include this approach in our experiments as it allows us to test whether increasing identifier diversity improves the GR's ability to represent document semantics accurately.

% This section ...
%This section provides an overview of the information retrieval paradigms relevant to this work, from traditional sparse methods to modern dense and generative models. Furthermore, we discuss recent works studying the differences and limitations of these approaches.

% short comment about sparse / bm25 results on LIMIT
% "Low-Rank Bottleneck" in more detail? sign-rank, etc.
% generative retrieval (general introduction / mechanism seq2seq, SEAL, MINDER, FM-index, docid designs (only if relevant!)

%We then formalize the low-rank bottleneck in dense retrieval, which is the theoretical bottleneck that the LIMIT benchmark is based on.
\section{Experimental Setup} \label{sec:experimental_setup}

In this section, we describe the evaluation setup for the GR models explained in Section~\ref{sec:preliminaries}.
We provide more details on the LIMIT dataset, metrics and baselines for our evaluation. 
We also introduce several modifications to the original SEAL and MINDER configurations that helped us boost their performance on LIMIT.
All experiments were conducted on a single NVIDIA H100L-94C GPU with 94 GB of VRAM.
Our code and experimental scripts for preprocessing and evaluation are publicly available at \url{https://anonymous.4open.science/r/GR-on-LIMIT/}.

%To allow for comparison with prior work, we adopt Recall@k as evaluation metric. As the benchmark's queries each have exactly two relevant documents, Recall@2 is the primary metric for task success. We also report Recall@10 and Recall@100 for a more complete analysis. 

\subsection{LIMIT Dataset}
\label{sec:limit}
The original LIMIT dataset was generated by prompting Gemini 2.5 Pro for 1,850 unique objects (items) that a person may like, such as ``Hawaiian pizza'', ``sports cars'', etc.
Then the authors sample unique subsets of 50 such items to create 50k documents, where each document is a single sentence generated using the \textit{document template}: ``<person> \textit{likes} <item1>, <item2>, <item3> ... <item50>''.
By sampling a unique name for every document, LIMIT generates sentences that express all preferences of a particular person, such as, ``John Doe \textit{likes} Apples, Quokkas, ...''
The sampling script is designed such that every item is liked by exactly two people.

The authors then generate 1,000 queries by picking one of the original items and filling the \textit{question template}: such as ``\textit{Who likes} Quokkas?''
Since every such query has exactly two relevant documents by design, the total number of relevant documents is 46 since it is the minimum number of documents relevant for 1,000 queries: $\binom{46}{2} = 1035$. Statistics on the dataset are given in Table~\ref{tab:LIMIT_statistics}.

\subsection{Metrics}
Recall@k is our primary evaluation metric in all our experiments.
Since every query in LIMIT is known to have exactly two relevant documents, Recall@2, which, in this setting, is the same as Precision@2, is the primary indicator of task success. We additionally provide Recall@10 and Recall@100 as an auxiliary metric for a complete analysis of the ranking quality.
% We report Recall@4 on LIMIT-H and LIMIT-HS as we extend the original LIMIT with up to 2 hard negative samples per relevant document.

\begin{table}
  \caption{Statistics of the original LIMIT dataset.}
  \label{tab:LIMIT_statistics}
  \begin{tabular}{ccccc}
    \toprule
    \#Docs & \#Queries & \#Relevant Docs & Avg Doc Length & \#Qrels \\
    \midrule
    50,000 & 1,000 & 46 & 598 & 2,000 \\
  \bottomrule
\end{tabular}
\end{table}

% Baselines
\subsection{Baselines}
We adopt the original set of baseline models from \citet{limit} that provides comprehensive results for BM25 and state-of-the-art single-vector and multi-vector DR models: 
\begin{itemize}
    \item \textbf{BM25}, a classic sparse IR method relying on exact term matching;
    \item \textbf{E5-Mistral 7B}, a instruction-tuned model based on Mistral;
    \item \textbf{Snowflake Arctic L}, a large-scale model optimized for retrieval;
    \item \textbf{GritLM 7B}, a model capable of generation and representation;
    \item \textbf{Promptriever Llama3 8B}, which is trained with diverse instructions to enable prompting like a language model;
    \item \textbf{Qwen3 Embed}, a SoA embedding model from the Qwen family;
    \item \textbf{Gemini Embed}, Google's text embedding model;
    \item \textbf{GTE-ModernColBERT}, a bi-encoder which utilizes late interaction and multiple vectors per document to capture fine-grained relevance.
\end{itemize}

% SEAL configuration
\subsection{SEAL Configurations} \label{subsection:seal_configs}
%As the first of our GR models under investigation, we include FM-index constrained ngram-based retrieval, first proposed in SEAL, because the docid design is especially well suited to represent the documents. For example, the document ``John Doe likes Quokkas, Apples, ...'' yields the ngrams ``Quokkas'', ``Apples'', among others. This gives SEAL a definitive target when queried  ``Who likes <item>?'', a pattern that all LIMIT queries follow. This ability to pinpoint and match exact, arbitrary substrings makes it an ideal GR method to contrast with embedding-based methods. Furthermore, SEAL's successful application to challenging IR benchmarks like NQ~\cite{natural_questions} and KILT~\cite{kilt} demonstrates that it is a robust retrieval method that generalizes to larger-scale, non-synthetic data.

%This reliance on ngrams accurately representing the document content is central to our subsequent investigation.
%\paragraph{Configurations}
We use the pre-trained LM checkpoint based on BART-large from the original paper, which was fine-tuned on the Natural Questions (NQ) dataset~\cite{bart}.
We did not fine-tune on the LIMIT dataset to match the zero-shot setting.
% This ensures that the model's performance is a result of its inherent architectural properties and pre-trained knowledge, rather than overfitting to the specific task.
The hyperparameters of the original model are also unchanged (beam width of 15, the score exponent ($\alpha$) is 2, repetition penalty ($\beta$) is 0.8).

We used the official implementation of SEAL~\footnote{\url{https://github.com/facebookresearch/SEAL}} with the default configuration and also introduced a minor modification to its original scoring procedure (BEAM):

\begin{itemize}
    \item \textbf{SEAL}: In its default configuration, SEAL does not follow the classic beam search algorithm (see Section~\ref{sec:seal} for details on how SEAL works).
    \item \textbf{SEAL (BEAM)}: We modified SEAL to use the classic beam search algorithm instead, which allowed it to keep only the highest scoring ngrams as opposed to the more conservative original SEAL implementation.
\end{itemize}
    % keeps because we observed that the original implementation often led to generic ngrams outweighing more informative ngrams.
    % Thus, we also evaluate SEAL with the classical beam search algorithm, keeping only the highest scoring ngrams.
    %\item \textbf{}: In the default configuration SEAL uses intersective scoring, taking only the highest scoring of a set of overlapping ngrams into account and ignoring the others to avoid multi-scoring of the same substrings. However, we observe that, in the case of LIMIT, this leads to the problem that ambiguous ngrams sometimes block more definitive ones. For example, ``Science'' may be scored higher than and therefore block ``Science Fiction'', although the latter would definitively identify all relevant documents.
    %\item Lastly, we evaluate a configuration with both modifications. That is, only consider highest-scoring ngrams and allow intersections.

% MINDER configurations
\subsection{MINDER Configurations}
%We include MINDER to evaluate whether diversifying document identifiers enhances robustness against the semantic limitations of ngrams. MINDER extends SEAL by introducing a \textit{multi-view} identifier approach, augmenting the index with synthetic pseudo-queries generated by a sequence-to-sequence model, and the title of the document. However, since the LIMIT documents do not include titles, we restrict our analysis to ngrams and pseudo-queries.

Similar to the SEAL setup, we used the original model checkpoint~\footnote{\url{https://github.com/liyongqi67/MINDER}} provided by \citet{minder}, which was also fine-tuned on NQ from BART large.
We used the default hyperparameters, which are the same as in SEAL.
Since the authors did not publicly release their question generation model, we used docT5query instead~\cite{doct5query}, which is a popular LM used for document expansion. 

We evaluate five different configurations of MINDER to ablate the impact of additional docids (views) and scoring mechanisms:

\begin{itemize}
    \item \textbf{NG}: Only ngrams as docids with default SEAL decoding.
    \item \textbf{NG; BEAM}: Only ngrams as docids with our SEAL (BEAM) decoding.
    \item \textbf{NG+DT5Q}: The default MINDER configuration, with ngrams and generated pseudo-queries under default SEAL decoding.
    %\item \textbf{DT5Q; Def.}: Here we isolate the efficiency of pseudo-queries as document identifier, more specifically, the ones generated by docT5query.
    \item \textbf{NG+DT5Q; BEAM}: Ngrams and generated pseudo-queries with our SEAL (BEAM) decoding.
    \item \textbf{Oracle}: Ngrams are generated with our SEAL (BEAM) decoding and perfect pseudo-queries as the skyline for the query generation performance. We generate 50 pseudo-queries for each document using the same \textit{template} as the actual queries in LIMIT: ``\textit{Who likes} <item>?''. This oracle helps us to determine the upper-bound for the query generation performance instead of the docT5query results.
\end{itemize}

\section{Results on LIMIT}
\label{sec:results_on_limit}
%In this study, we empirically investigate the claim that GR can bypass the compression limitations that constrain dense retrieval models.

%Ranking in GR is a generation task that models the conditional probability $P(\text{docid}_i|q)$ of a document identifier $\text{docid}_i$ given query $q$, typically using the transformer architecture. Thereby, relevance information is not compressed into a fixed-dimensional vector space as in embedding models, but encoded directly into the model's parameters $\theta$, which are the weights of the transformer. Since transformers, given sufficient capacity, are universal sequence approximators~\cite{transformers_approximators, universal_approximators}, they possess the architectural ability to approximate any high rank relevance matrix with arbitrarily small error.

In this section, we present the results of our experimental evaluation on the original LIMIT benchmark.
Table~\ref{tab:results_limit} compares the performance of our GR models, and their modifications, with the DR and BM25 results reported by \citet{limit}.
% to encode relevance structures that exceed the capacity of fixed-dimensional embeddings.
% Our objective is to verify if the theoretical architectural advantages of GR translate to practical success. 

% We present the results of our evaluation on the full 50k-document LIMIT benchmark in Table~\ref{tab:results_limit}. Baseline results are taken from~\citet{limit}.

\begin{table}
  \caption{Retrieval results on the original LIMIT dataset (50k documents). Baseline results (*) are from~\citet{limit}. Bold indicates the best overall performance.}
  \label{tab:results_limit}
  \begin{tabular}{l rrr}
    \toprule
    Model & R@2 & R@10 & R@100 \\
    \midrule
    * E5-Mistral 7B (4096) & 0.013 & 0.022 & 0.083 \\
    * GritLM 7B (4096) & 0.024 & 0.041 & 0.129 \\
    * Promptriever Llama3 8B (4096) & 0.030 & 0.068 & 0.189 \\
    * Qwen3 Embed (4096) & 0.008 & 0.018 & 0.048 \\
    * Gemini Embed (3072) & 0.016 & 0.035 & 0.100 \\
    * GTE-ModernColBERT & 0.231 & 0.346 & 0.548 \\
    * Snowflake Arctic L V2 (1024) & 0.004 & 0.008 & 0.033 \\
    * BM25 & 0.857 & 0.904 & 0.936 \\
    \hline
    SEAL & 0.791 & {\textbf{0.999}} & {\textbf{1.000}} \\
    SEAL (BEAM) & 0.917 & 0.982 & 0.991 \\
    MINDER (NG) & 0.858 & 0.994 & \textbf{1.000} \\
    MINDER (NG+DT5Q) & 0.858 & 0.994 & \textbf{1.000} \\
    MINDER (NG; BEAM) & \textbf{0.958} & 0.983 & 0.990 \\
    MINDER (NG+DT5Q; BEAM) & \textbf{0.958} & 0.983 & 0.990 \\
    \hline
    MINDER (Oracle) & 0.988 & 0.994 & 0.997 \\
  \bottomrule
\end{tabular}
\end{table}

\paragraph{Comparison with Baselines} 
% As reported by~\citet{limit}, single-vector embedding models fail to capture the high-rank relevance ($\leq$ 0.030 R@2). The multi-vector GTE-ModernColBERT performs substantially better (0.231 R@2), but still falls far behind the sparse lexical model BM25 (0.857 R@2). 
The GR models demonstrate superior capacity for high-rank retrieval with SEAL achieving up to 0.917 R@2 and MINDER offering further improvements to 0.988 R@2, which is a substantial gain over all the baselines including BM25, which was the best model reported in \citet{limit}. 
Thereby, the GR models (MINDER and SEAL) set the new state-of-the-art results on the LIMIT benchmark outperforming the DR models by a large margin. 

\paragraph{Impact of Decoding Strategy}
We also observe that our modification to the standard SEAL's decoding procedure, greatly improves the model performance for both SEAL and MINDER.
The default SEAL configuration achieves perfect Recall@100 ($1.000$) but lower precision at the top ranks ($0.791$ R@2) compared to our beam search configuration ($0.917$ R@2). Similarly, for MINDER, we improve R@2 from $0.858$ for the default SEAL to $0.958$ with beam search.

\paragraph{MINDER and Pseudo-Queries} 
% MINDER consistently outperforms SEAL in comparable settings ($0.958$ vs. $0.917$ R@2). 
Generating pseudo-queries with docT5query (DT5Q) yields no improvement over ngrams only configuration.
Interestingly, even in the oracle setting, with the perfect pseudo-queries that exactly match the input queries, the performance of MINDER is at $0.988$ R@2.
We manually checked the results to confirm that MINDER does not copy the input queries to be docids, which would have trivially solved this task to $1$ R@2.
This observation suggests that the additional query-to-query generation introduces another potential point of failure and degrades the GR performance even when the document expansion step succeeds.

% \section{Harder Benchmark Construction} 
% In the previous section we confirm that GR is capable of overcoming the compression limitations of dense retrieval. However, in this section we argue that the original LIMIT benchmark does not sufficiently evaluate the capability to represent semantics. Notably, the high performance of BM25 on LIMIT (see Section~\ref{sec:results_on_limit}) suggests that the task is reduced to combinatorial pattern matching and does not require semantic disambiguation, which is the main reason for the generational paradigm shift from sparse to dense retrieval. In other words, a model can achieve high recall on LIMIT without necessarily encoding the contextual meaning required to distinguish between semantically distinct usages of those terms.

% We argue that an ideal retrieval system is capable of overcoming compression limitations while capturing semantics accurately. Motivated by this aim, we introduce benchmark variants LIMIT-H and LIMIT-HS, that are explicitly designed to contain the exact identifiers a model is trained to generate (e.g., ngrams) but are irrelevant to the query while maintaining the query-document relevance patterns that expose compression limitations. For ngrams, this comes down to breaking the one-to-one mapping between terms and documents.

\section{New LIMIT-H and LIMIT-HS Benchmarks}
Our results, presented in Section~\ref{sec:results_on_limit}, confirm that GR is capable of overcoming the vector bottleneck of DR.
However, it is clear that the synthetic LIMIT benchmark does not reflect all the complexities of a real-world IR dataset.
In particular, LIMIT was not designed to evaluate the capacity of the model to represent fine-grained semantic relationships between documents, which are arguably more common in IR in practice than strict logical constraints.

The main simplification in LIMIT, which is due to its rule-based design, is that every item (e.g. ``Quokkas'') can be used to uniquely identify the complete set of relevant documents.
This pattern explains why GR models succeed on LIMIT as long as they are able to generate those items as docids.

We propose an extension to LIMIT, which we call LIMIT-H (for LIMIT Hard), that makes those simple identifiers more ambiguous and contextually dependent.
In LIMIT-H, we deliberately introduce such \textit{identifier ambiguity} by generating new documents that are very similar to the original relevant documents but are not relevant to the corresponding queries (see Figure~\ref{fig:ambiguous_docid_retrieval} for an example).
% Those new documents (hard negative samples) share the same ngrams that are generated by GR from the query as docids but diverge semantically s.t. the document is no longer relevant to the query . 

Our dataset generation pipeline proceeds in three stages:

\paragraph{1. Generating Negative Documents}
For each of the 46 relevant documents from the original LIMIT dataset, we generate a corresponding negative sample by prompting the DeepSeek-R1 70B model from Ollama with its default inference parameters.
Our few-shot prompt\footnote{\url{https://anonymous.4open.science/r/GR-on-LIMIT/prompts.py}} takes as input an original relevant document and instructs the model to generate a similar sentence mentioning the same 50 items but reversing its semantics, such that the original preference statements can be no longer implied from the generated text (see a sample negative \textbf{D1*} in Figure~\ref{fig:ambiguous_docid_retrieval}).
% This new document is designed to share ngrams that are likely to be generated by the query, such as mentioning the same items with the same capitalization, but has a different meaning overall, such as the example in Figure~\ref{fig:ambiguous_docid_retrieval}.
At the end of the prompt, we provide one example of correct generation and one example of incorrect generation with a brief summary reasoning why the generated negative is considered incorrect.
We also introduce several explicit constraints in the prompt to specify quality of the generated samples:
\begin{enumerate}
    \item the negative must start with the same three words as the original document (e.g., ``John Doe likes'');
    \item the negative must include every item from the original document and preserve the exact spelling and capitalization as in the original document;
    \item the negative must have distinct semantics from the original document; it cannot contain duplicates or paraphrases of the original preference statements.
\end{enumerate}

To ensure quality of the generated samples, we prepared a validation script that checks the first two constraints automatically and ensures that the generated sample is not the exact duplicate of the original for the third constraint.
If any of the constraints were violated, we automatically rerun the generation using the same input document, which results in a different output every time due to probabilistic decoding (default Ollama settings).

The most frequent generation error, which we could efficiently catch with our automated verification script, was that some of the items mentioned in the original document were missing from the generated document.
To efficiently revise the generated sample, we designed another ``completion'' prompt, which given the generated sample and the set of missing items, asks the model to add the missing items to the generated sample.
Thereby, we iteratively perform this data generation process with the automated validation check that triggers regeneration for the same input document until all checks are passed for all the input documents.

This automated generation task took $\sim$2--3 hours for all 46 relevant documents from LIMIT.
Afterwards, we manually checked that all the generated documents satisfy the third (semantic) constraint.
Fewer than 5 documents violated the semantic constraint and we ran the generation task on this subset again until the manual check was passed for all the input documents.
We noticed that the LM generated diverse phrases that successfully modify the original items within the same document but may repeat same or similar phrases across the generated documents.

\paragraph{2. LIMIT Benchmark Extension}
LIMIT-H contains the same documents, queries and qrels as the original LIMIT dataset, and only adds 46 new documents that we generated as explained in 1.
Therefore, the total number of documents in LIMIT-H is increased to 50046.
We release only those 46 additional documents and a script that automatically downloads the original LIMIT dataset and integrates it into LIMIT-H.

This is the only change we made to the original LIMIT benchmark.
However, this extension implies that there is one hard negative for each relevant document from LIMIT, and a cluster of 4 lexically similar documents for each query, which makes the task harder than the original LIMIT.

\paragraph{Generating Oracle Pseudo-Queries for Negative Documents}
For MINDER we need to generate oracle pseudo-queries and add them to the FM-index.
The oracle pseudo-queries for the original documents were deterministically constructed using the same template as for the original queries in LIMIT and their item set.
Since we used a LM to generate negative documents, the original pattern for the oracle pseudo-queries no longer holds.
Instead of using the pre-trained docT5query model for expanding the negative documents, we use the same DeepSeek-R1 70B model with a few-shot prompt, which is specifically optimized for LIMIT: generate questions starting with ``Who likes ''.
We consider this as a good oracle for our purposes but it is not designed to generalize to other datasets.

\paragraph{LIMIT-HS}
LIMIT-HS (for LIMIT Hard-Scaled) is a further extension of LIMIT-H, in which we trivially duplicate the 46 generated negative documents from LIMIT-H.
This extension results in two hard negative samples per relevant document and a cluster of 6 lexically similar documents for each query from LIMIT.
Note, that this simple approach allows us to maintain the same set of ngrams and oracle pseudo-queries in the FM-index.
Thereby, we control for the only variable condition in the LIMIT-HS evaluation: the trivial increase in the number of negative instances.
% LIMIT-HS is intended to measure the decline in model performance as identifier ambiguity increases further.
% We duplicate the 46 negative documents generated for LIMIT-H and assign them new, unique document IDs.
% This results in a 2:1 ratio of negatives to positives, while not introducing additional decoding targets, i.e., 

% ngrams and pseudo-queries are the same as in LIMIT-H, only the number of negatives is increased.

% documents. This creates a 1:1 ratio of relevant to irrelevant-but-lexically-matching documents for each query, effectively turning unique ngrams into ambiguous ones.

\paragraph{Introducing Variance}
The original LIMIT dataset is very minimalistic and homogenic since it uses a single template for rule-based natural language generation (see Section~\ref{sec:limit} for the LIMIT generation approach).
The constrained decoding via the FM-index used by SEAL and MINDER is not probabilistic and leads to the same docids for the same inputs.
To better evaluate variance in the model results but also stay close to the original LIMIT approach, we created two additional versions of LIMIT-H with alternative documents that describe other relations that can hold between a set of people and a set of items rather than preference.
To achieve this, we simply substitute the verb ``likes'' in the original LIMIT templates for documents and queries with another two verbs: ``owns'' and ``touches''.
These alternative relations were selected to also cover more diverse semantic categories: ``owns'' implies a relation of possession, which is more specific than the preference relation, and ``touches'' denotes a physical action rather than a nominal relation between a person and items.
Moreover, ``touches'' is selected to ensure that the relation is not transitive: $A$ touches $B$ and $B$ touches $C$ does not imply $A$ touches $C$.
% , thereby eliminating the semantic ambiguity found in ``likes'', where an affinity for a subject (e.g., a book about Quokkas) might erroneously suggest an affinity for the object itself (Quokkas).
We then perform the same dataset generation procedure as for ``likes'' to generate alternative negative documents and their pseudo-queries using the same scripts and prompts.
Those documents and their pseudo-queries are also sufficient to extend both LIMIT and LIMIT-HS since they contain the same documents as LIMIT-H, where LIMIT is the subset and LIMIT-HS is a superset of LIMIT-H.

\section{Results on LIMIT-H and LIMIT-HS} \label{sec:results_limit-h-hs}
\begin{figure}
  \centering
  \includegraphics[width=\linewidth]{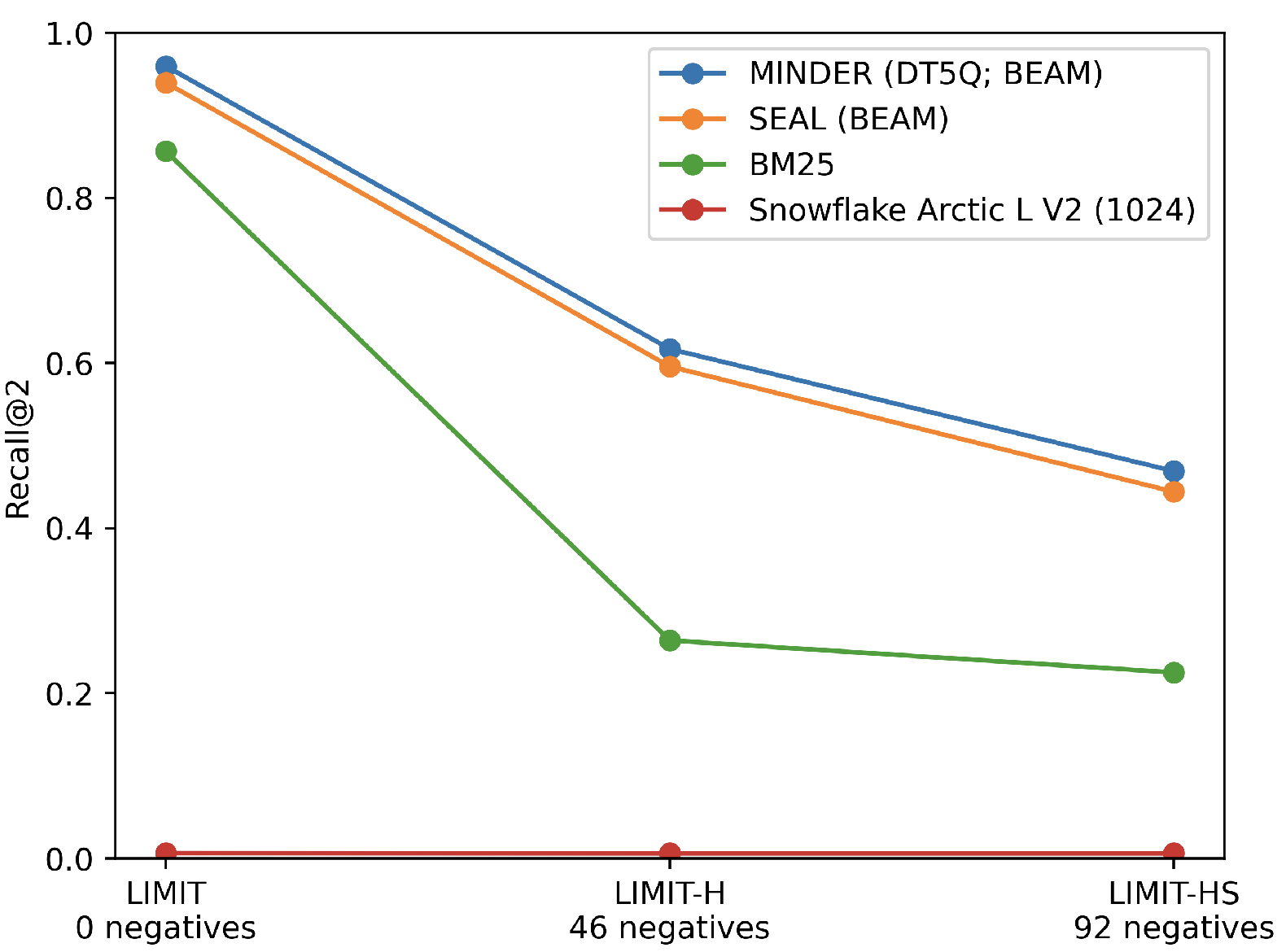} 
  \caption{Comparison of Recall@2 performance across three increasing levels of difficulty.}
  \label{fig:hard_scaling}
  \Description{Recall@2 drops strongly when introducing negative documents, demonstrating that docid ambiguity is a real vulnerability of SEAL and performance degradation scales with ambiguity.}
\end{figure}

Table~\ref{tab:semantic_robustness} shows the results on new LIMIT-H and LIMIT-HS variants of the LIMIT benchmark.
We computed the metric for each the three variants of the datasets independently (with ``likes'', ``owns'' and ``touches'' relations) and measured the mean and standard deviation across the results.
% We evaluate only the 
% Our results reveal distinct failure modes across retrieval paradigms, and allow for a more comprehensive comparison of these methods.
% The results reveal three distinct failure modes across retrieval paradigms.
Figure~\ref{fig:hard_scaling} compares four models with their best-performing configurations on all LIMIT variants.

\begin{table*}
  \caption{Mean and standard deviation of R@2 for three variants (``likes'', ``owns'', ``touches'') each of LIMIT, LIMIT-H and LIMIT-HS. Bold indicates the best overall performance (excluding the oracle configuration). $\dagger$ indicates statistically significant improvement ($p < 0.05$) over the strong BM25 baseline (paired t-test).}
  \label{tab:semantic_robustness}
  \begin{tabular}{l ccc ccc}
    \toprule
    Model & \multicolumn{3}{c}{Recall@2} & \multicolumn{3}{c}{Recall@4} \\
          & LIMIT & LIMIT-H & LIMIT-HS   & LIMIT & LIMIT-H & LIMIT-HS \\
    \midrule
    BM25                          & 0.857 $\pm$ 0.000 & 0.249 $\pm$ 0.049 & 0.208 $\pm$ 0.067 & 0.893 $\pm$ 0.000 & 0.717 $\pm$ 0.083 & 0.462 $\pm$ 0.160\\
    %Snowflake Arctic L V2         & 0.005 $\pm$ 0.003 & 0.005 $\pm$ 0.003 & 0.005 $\pm$ 0.003 & 0.007 $\pm$ 0.005 & 0.007 $\pm$ 0.005 & 0.007 $\pm$ 0.004 \\
    \midrule
    SEAL                          & 0.851 $\pm$ 0.052 & 0.000 $\pm$ 0.000 & 0.000 $\pm$ 0.000 & {\textbf{0.975}}\rlap{$^{\dagger}$} $\pm$ 0.011 & 0.000 $\pm$ 0.001 & 0.000 $\pm$ 0.000 \\
    SEAL (BEAM)                   & {0.947}\rlap{$^{\dagger}$} $\pm$ 0.026 & {0.589}\rlap{$^{\dagger}$} $\pm$ 0.012 & {0.433}\rlap{$^{\dagger}$} $\pm$ 0.021 & {0.965}\rlap{$^{\dagger}$} $\pm$ 0.013 & {0.951}\rlap{$^{\dagger}$} $\pm$ 0.022 & {0.823}\rlap{$^{\dagger}$} $\pm$ 0.016 \\
    MINDER (NG+DT5Q)              & {0.910}\rlap{$^{\dagger}$} $\pm$ 0.050 & 0.015 $\pm$ 0.020 & 0.001 $\pm$ 0.001 & {0.970}\rlap{$^{\dagger}$} $\pm$ 0.015 & 0.088 $\pm$ 0.180 & 0.000 $\pm$ 0.000 \\
    MINDER (NG; BEAM)             & {0.945}\rlap{$^{\dagger}$} $\pm$ 0.028 & {0.595}\rlap{$^{\dagger}$} $\pm$ 0.041 & {0.443}\rlap{$^{\dagger}$} $\pm$ 0.048 & {0.965}\rlap{$^{\dagger}$} $\pm$ 0.013 & {0.949}\rlap{$^{\dagger}$} $\pm$ 0.030 & {0.825}\rlap{$^{\dagger}$} $\pm$ 0.047 \\
    MINDER (NG+DT5Q; BEAM)        & {\textbf{0.954}}\rlap{$^{\dagger}$} $\pm$ 0.010 & {\textbf{0.602}}\rlap{$^{\dagger}$} $\pm$ 0.027 & {\textbf{0.450}}\rlap{$^{\dagger}$} $\pm$ 0.034 & {0.969}\rlap{$^{\dagger}$} $\pm$ 0.007 & {\textbf{0.961}}\rlap{$^{\dagger}$} $\pm$ 0.010 & {\textbf{{0.834}}}\rlap{$^{\dagger}$} $\pm$ 0.026 \\
    \midrule 
    MINDER (Oracle)               & {0.986}\rlap{$^{\dagger}$} $\pm$ 0.003 & {0.655}\rlap{$^{\dagger}$} $\pm$ 0.031 & {0.509}\rlap{$^{\dagger}$} $\pm$ 0.056 & {0.991}\rlap{$^{\dagger}$} $\pm$ 0.001 & {0.988}\rlap{$^{\dagger}$} $\pm$ 0.001 & {0.874}\rlap{$^{\dagger}$} $\pm$ 0.021 \\
    \bottomrule
\end{tabular}
\end{table*}

\paragraph{Baseline Performance}
The performance of Snowflake Arctic L remains consistently low across all variants ($0.005 \pm 0.003$ R@2), showing no significant sensitivity to the added ambiguity. The model's failure is fundamental to the combinatorial nature of the LIMIT dataset itself (geometric limitations), rather than the introduced term ambiguity. The added difficulty of LIMIT-H/HS is effectively irrelevant because the model cannot solve the base task.
%DR baselines were excluded from Table~\ref{tab:semantic_robustness} due to near-zero performance on the easier LIMIT task

In contrast, BM25 performs strongly on the original benchmark (0.857 R@2) but much worse when introducing term-ambiguous negatives in LIMIT-H (0.249 R@2) and LIMIT-HS (0.208 R@2). Notably, R@2 on LIMIT-H falls below half of what it is on LIMIT, meaning that it is even worse than failing to distinguish: BM25 tends to rank negatives higher than relevant documents.

%In the hardened variants, BM25 retrieval fails much more frequently because it relies on the ``bag-of-words'' assumption, retrieving negatives with equal probability based only on term frequency.

\paragraph{Docid Ambiguity in Generative Retrieval}
%\paragraph{LM+FM Collapse} 
The default configurations for SEAL and MINDER collapse to near-zero recall (< 0.001 R@2) on the LIMIT-H and LIMIT-HS. 

Using classical beam search (BEAM) mitigates this collapse, allowing GR methods to substantially outperform BM25 and Snowflake Arctic L V2 on all benchmark variants. However, we still observe a substantial drop in recall. SEAL (BEAM) drops from $0.947$ (LIMIT) to $0.433$ (LIMIT-HS) R@2. MINDER (NG+DT5Q; BEAM) similarly drops from $0.954$ to $0.450$ R@2. Notably, MINDER improves retrieval overall, but does not offer more robustness to term-ambiguity. Even when perfect pseudo-queries (PQ) are assumed, MINDER fails to consistently identify negatives ($0.655$ and $0.509$ R@2 on LIMIT-H and LIMIT-HS, respectively).

\section{Error Analysis} \label{sec:error_analysis}
In this section, we introduce a novel quantitative framework specifically designed to measure how generated docids, in particular ngrams and pseudo-queries, contribute to the document ranking performance.
We compare the sets of docids generated for relevant and irrelevant documents.
% and their hard negative samples allow us to gain a more in-depth understanding of the DR models' performance.
% mechanisms behind successful and failed retrieval in the computational study in Sections~\ref{sec:results_on_limit} and \ref{sec:results_limit-h-hs}.

This error analysis framework helps us to answer the following questions:
% concrete research questions about how retrieval failure manifests in generative retrieval, in particular when faced with identifier ambiguity:
% we aim to answer the following questions:
\begin{itemize}
    \item [Q1] How does identifier ambiguity impact retrieval in SEAL and MINDER?
    \item [Q2] Why does SEAL default decoding fail on LIMIT-H?
    \item [Q3] Why the performance degrades further on LIMIT-HS?
    \item [Q4] Why the oracle pseudo-queries in MINDER do not improve the retrieval performance?
\end{itemize}

\subsection{Method}
To analyse error types, we focus on the top-4 retrieved documents ($D_q$) and the docids generated for those documents by our GR models.
Since each query $q$ in LIMIT-H contains two relevant documents and two hard negatives, top-4 is likely to capture all top scoring documents.
Our analysis is performed in three steps:
\begin{enumerate*}[]
\item partition rankings,
\item partition docids, and
\item aggregate scores.
\end{enumerate*}

 % We maintain this evaluation window across all benchmark variants to study model behavior as negatives are introduced.

\paragraph{1. Partition rankings.} $D_q$ consists of relevant ($R$) and irrelevant ($I$) subsets of documents. Based on the relevance of the top-2 documents, each ranking result falls into one of the four partitions (\textbf{groups}): $o \in \{RRXX, RIXX, IRXX, IIXX\}$. For instance, $RIXX$ denotes a ranking where the highest ranked document is relevant ($d_1 \in R$) and second highest ranked document is irrelevant ($d_2 \in I$), while the relevance of the remained documents ($X$) can be either.

\paragraph{2. Partition docids.} We refer to the set of all docids generated by the model for the top-4 retrieved documents as $N_q$.
Those docids can either:
\begin{itemize}
    \item appear \textit{only} in relevant documents ($n \in N_q^{R \backslash I}$);
    \item appear \textit{only} in irrelevant documents ($n \in N_q^{I \backslash R}$).
    \item appear in \textit{both} relevant and irrelevant documents ($n \in N_q^{R \cap I}$).
\end{itemize}
Thereby, all docid $n$ have associated \textbf{types} $x \in \{R \backslash I,\, I \backslash R,\, R \cap I\}$.
%  (ambiguous identifiers)

\paragraph{3. Aggregate scores.} Finally, for each docid $n$,  query $q$ and document $d$, we compute a score $c(n, q, d) = w(n, q)^\alpha \cdot \textit{cover}(n, K^{(d)})$ as per Equation~\ref{eq:seal-document-level-score}.
We aggregate those scores for every query $q$ and each docid type $x$ separately: 

\begin{definition}[Query-Level Score]
\begin{equation}
c(q, x) = \sum_{d \in D_q} \sum_{n \in N_q^{x}} c(n, q, d)
\end{equation}
\end{definition}

Then, we aggregate those individual query scores across the previously defined groups of ranking results by averaging the scores across all queries $q \in Q_o$:

\begin{definition}[Group-Level Mean Score]
\begin{equation}
c_{g}(o, x) = \frac{1}{|Q_o|} \sum_{q \in Q_o} c(q, x)
\end{equation}
\end{definition}

% Absolute scores can be misleading if the total score magnitude varies significantly between experimental settings. 
To compare across groups, we normalize the scores across types of docids to sum up to 1 for each group:

\begin{definition}[Group-Level Relative Score]
\begin{equation} \label{eq:relative_contribution}
c_{r}(o, x) = \frac{c(o, x)}{\sum_{y \in \{R \backslash I,\, I \backslash R,\, R \cap I\}}C(o, y)}
\end{equation}
\end{definition}

%Given a query $q_i$, SEAL applies BART to generate ngrams $N_i = \textrm{BART(Q)}$. For each query, we group these ngrams by their appearance in the four highest-scoring documents $D_i^4$, according to the respective documents' relevance. More specifically, as in the original SEAL paper, we consider only the highest scoring ngram whenever multiple ngrams intersect. We denote the ngrams as $N_i^R$, appearing \textit{only} in relevant documents, and $N_i^I$ for ngrams in irrelevant documents. Ngrams that are shared between \textit{both}, relevant and irrelevant, documents, are also grouped together as $N_i^S$.

%Note that this is a partition of all generated ngrams that appear at least once in the top-4 retrieved documents. The contribution of an ngram $n$ to the document-level score $W(d,q)$ is $w(n, q)^\alpha * \textit{cover}(n, K^{(d)})$, as shown in Equation~\ref{eq:seal-document-level-score}. For shorthand, we denote the contribution of an ngram as $c(n, q, K^{(d)})$. Finally, what we compare is the relative contribution (Equation~\ref{eq:relative_contribution}) of ngrams in relevant, irrelevant and both types of documents to the total score over all four documents.

\subsection{Error Analysis Results}

\begin{table}
\caption{Error analysis of SEAL in the default configuration. The model fails to retrieve any relevant documents at the top-2 ranks for LIMIT-H and LIMIT-HS, resulting in no score being computed (-).}\label{tab:ngram-analysis-LM+FM} %Empty categories ($Q=\emptyset$) are omitted.
\begin{tabular}{lc rccc}
\toprule
Dataset & $o$ & $|Q_o|$ & $c_{r}(o, R \backslash I)$ & $c_{r}(o, I \backslash R)$ & $c_{r}(o, R \cap I)$ \\
\midrule
LIMIT & \verb|RRXX| &652 &0.17 &0.24 &0.59 \\
& \verb|RIXX| &139 &0.10 &0.30 &0.60 \\
& \verb|IRXX| &138 &0.10 &0.29 &0.61 \\
& \verb|IIXX| &71 &0.06 &0.46 &0.49 \\
\midrule
%LIMIT-H & & & & & \\
LIMIT-H & \verb|RRXX| & 0 & - & - & - \\
&\verb|RIXX| & 0 & - & - & - \\
&\verb|IRXX| & 0 & - & - & - \\
&\verb|IIXX| &1000 &0.00 &1.00 &0.00 \\
\midrule
%LIMIT-HS & & & & & \\
LIMIT-HS & \verb|RRXX| & 0 & - & - & - \\
&\verb|RIXX| & 0 & - & - & - \\
&\verb|IRXX| & 0 & - & - & - \\
& \verb|IIXX| &1000 &0.00 &1.00 &0.00 \\
\bottomrule
\end{tabular}
\end{table}

\begin{table}
\caption{Error analysis of SEAL in the classical beam search configuration (BEAM).}\label{tab:ngram-analysis-BEAM}
\begin{tabular}{lc cccc}
\toprule
Dataset & $o$ & $|Q_o|$ & $c_{r}(o, R \backslash I)$ & $c_{r}(o, I \backslash R)$ & $c_{r}(o, R \cap I)$ \\
\midrule
%LIMIT & & & & & \\
LIMIT & \verb|RRXX| &908 &0.63 &0.27 &0.10 \\
& \verb|RIXX| &5 &0.43 &0.06 &0.51 \\
& \verb|IRXX| &13 &0.34 &0.39 &0.27 \\
& \verb|IIXX| &74 &0.13 &0.86 &0.01 \\
\midrule
%LIMIT-H & & & & & \\
LIMIT-H & \verb|RRXX| &316 &0.00 &0.00 &1.00 \\
& \verb|RIXX| &232 &0.00 &0.00 &1.00 \\
& \verb|IRXX| &323 &0.00 &0.01 &0.99 \\
& \verb|IIXX| &129 &0.02 &0.63 &0.36 \\
\midrule
%LIMIT-HS & & & & & \\
LIMIT-HS & \verb|RRXX| &316 &0.00 &0.00 &1.00 \\
& \verb|RIXX| &229 &0.00 &0.00 &1.00 \\
& \verb|IRXX| &41 &0.00 &0.04 &0.96 \\
& \verb|IIXX| &414 &0.01 &0.41 &0.58 \\
\bottomrule
\end{tabular}
\end{table}

In the following, we answer the research questions introduced in the beginning of this Section.

\paragraph{Q1: Retrieval under docid ambiguity}
Although recall metrics on LIMIT-H and LIMIT-HS (up to 0.655 and 0.509 R@2) suggest that generative retrieval is partially successful, the underlying retrieval mechanism is fundamentally broken.
As Tables~\ref{tab:ngram-analysis-BEAM} and \ref{tab:identifier-analysis-MINDER-NG+PQ} reveal, SEAL fails to generate \textit{any} ngrams unique to relevant documents in these settings.
Consequently, the model is no longer performing targeted retrieval. Instead, retrieval success is reduced to a combination of random chance, the avoidance of negative-only identifiers, and scoring heuristics that happen to favor properties of (incidentally) relevant documents (e.g., length, identifier coverage).
This collapse in discriminative capability is confirmed by the score margins: the gap between the rank-1 and rank-4 documents shrinks substantially. For SEAL (BEAM), on average by 92\% from 61.22 in LIMIT to just 4.95 in LIMIT-H.
For MINDER (NG+PQ), on average by 99\% from 15.87 to 0.19. This indicates that the models have lost the ability to distinguish relevant documents from noise.

\paragraph{Q2: Noise reduction via beam search}
The ngram analysis in Table~\ref{tab:ngram-analysis-LM+FM} demonstrates the effect of the default configuration accumulating score from all partially decoded ngrams, rather than restricting scoring to the top candidates.
This approach results in a high volume of generated ngrams in the ambiguous benchmark variants: 4.4 / 50.5 / 44.1 ngrams
per document on LIMIT / -H / -H using default scoring, and 2.3 /
1.5 / 1.5 ngrams per document with the BEAM configuration. More importantly, the additional ngrams occur almost exclusively in negative documents, dominating the score and resulting in complete failure (1000/1000).

\paragraph{Q3: Further performance degradation caused by displacement}
When comparing results for LIMIT-H and LIMIT-HS, we can see a further drop in R@2. Results in Table~\ref{tab:ngram-analysis-BEAM} suggest that this is mostly due to the further displacement of rank-2 relevant documents. This shows that some documents are more vulnerable to being displaced by negatives, and are ranked even lower when more of the same negatives are present. Simultaneously, there are relevant documents that could not be displaced by the corresponding negative in LIMIT-H, and consequently, cannot be replaced by an additional copy of the negative.

\paragraph{Q4: Pseudo-queries in MINDER fail to restore unique relevance signals}
We hypothesized that the inclusion of pseudo-queries would enrich the set of identifiers unique to relevant documents ($R \backslash I$), due to their ability to capture concepts and rely less on memorization than ngrams~\cite{zero_shot_with_pseudo_query}. The results in Table~\ref{tab:identifier-analysis-MINDER-NG+PQ} contradict this expectation. In both LIMIT-H and LIMIT-HS, the relative contribution of unique relevant identifiers $C_{r}(R \backslash I)$ remains at 0.00 across all error groups. This suggests that either (1) the initial pseudo-queries generation or the (2) scoring and decoding process, particularly early pruning, are the reason uniquely relevant pseudo-queries are not generated in our experiments.

Investigating pseudo-queries (Table~\ref{tab:query_comparison}) reveals that the PQ setting successfully generates document-specific questions that capture the meaning. This allows us to rule out generation quality as the root cause. Instead, the failure points conclusively to the scoring and decoding process: the retrieval mechanism prunes these unique, informative pseudo-queries during beam search.

\begin{table}
\caption{Error analysis of MINDER under the (BEAM, NQ+PQ) configuration, which employs manually constructed pseudo-queries for the original documents and few-shot LLM generated queries for negative examples.}\label{tab:identifier-analysis-MINDER-NG+PQ}
\begin{tabular}{lc cccc}
\toprule
Dataset & $o$ & $|Q_o|$ & $c_{r}(o, R \backslash I)$ & $c_{r}(o, I \backslash R)$ & $c_{r}(o, R \cap I)$ \\
\midrule
LIMIT & \verb|RRXX| &985 &0.82 &0.02 &0.17 \\
&\verb|RIXX| &2 &0.00 &0.00 &1.00 \\
&\verb|IRXX| &3 &0.17 &0.18 &0.64 \\
&\verb|IIXX| &10 &0.00 &0.91 &0.09 \\
\midrule
LIMIT-H &\verb|RRXX| &444 &0.00 &0.00 &1.00 \\
&\verb|RIXX| &230 &0.00 &0.00 &1.00 \\
&\verb|IRXX| &261 &0.00 &0.00 &1.00 \\
&\verb|IIXX| &65 &0.00 &0.18 &0.82 \\
\midrule
LIMIT-HS &\verb|RRXX| &445 &0.00 &0.00 &1.00 \\
&\verb|RIXX| &230 &0.00 &0.00 &1.00 \\
&\verb|IRXX| &21 &0.00 &0.00 &1.00 \\
&\verb|IIXX| &304 &0.00 &0.34 &0.66 \\
\bottomrule
\end{tabular}
\end{table}

\begin{comment}
\begin{table}\label{tab:query_comparison}
  \caption{Qualitative comparison of generated identifiers in docT5query (DT5Q) and oracle (PQ) setting.}
  \label{tab:semantic_robustness}
  \begin{tabular}{|p{0.1\linewidth}|p{0.4\linewidth}|p{0.4\linewidth}|}
    \toprule
     & Positive Document & Negative Document \\
    \midrule
    Doc.
    & 
    Olinda Posso likes Bagels, Hot Chocolate, Pumpkin Seeds, The Industrial Revolution, ...
    &
    Olinda Posso likes to imagine herself as the master architect of a surreal utopia, where Bagels levitate above Hot Chocolate rivers, ...
    \\
    \hline
    DT5Q
    &
    'who is olinda', 'olinda posesso likes what', 'what food brand does olinda poseo like', 'what songs did olinda posso sing', ... 
    &
    'what is the industrial revolution', 'what is the world of urban exploration', 'what are the industrial development and urban exploration theories', ... 
    \\
    \hline
    PQ
    &
    'Who likes Bagels?', 'Who likes Hot Chocolate?', 'Who likes Pumpkin Seeds?', ...
    &
    'Who likes to imagine herself as the master architect of a surreal utopia? ', 'Who likes Bagels levitating above Hot Chocolate rivers? ', ...
    \\
    \bottomrule
\end{tabular}
\end{table}
\end{comment}

% Identifiers in the Oracle setting using manually crafted pseudo-queries for original documents and few-shot LLM generated queries for negatives.

\begin{table}
  \caption{Sample documents with oracle pseudo-queries from LIMIT-H for the original query ``Who likes Hot Chocolate?''. Given this query, MINDER generates only the ngram: ``Hot Chocolate'', and no pseudo-queries scored above zero. }
  \label{tab:query_comparison}
  \begin{tabular}{ p{0.18\linewidth} p{0.355\linewidth} p{0.365\linewidth} }
    \toprule
    \textbf{Type} & \textbf{Document ngrams} & \textbf{Pseudo-queries} \\
    \midrule
    Original (Relevant) & Olinda Posso likes Bagels, \textbf{Hot Chocolate}, Pumpkin Seeds, The Industrial Revolution, ...
    &
    Who likes Bagels? Who likes \textbf{Hot Chocolate}? Who likes Pumpkin Seeds? ...
    \\
    \midrule
    Negative (Irrelevant) & Olinda Posso likes to imagine herself as the master architect of a surreal utopia, where Bagels levitate above \textbf{Hot Chocolate} rivers, ...
    &
    Who likes to imagine herself as the master architect of a surreal utopia? Who likes Bagels levitating above \textbf{Hot Chocolate} rivers? ...
    \\
    \bottomrule
\end{tabular}
\end{table}

\section{Discussion} \label{sec:discussion}

We briefly summarize and discuss our main findings.

\paragraph{GR outperforms DR}
% Our results highlight a fundamental distinction between the failure modes of dense and generative retrieval. 
The failure of DR models on the LIMIT benchmark originates from the vector bottleneck, where the fixed-dimensional embedding space lacks the capacity to linearly separate the combinatorial number of document subsets required by the LIMIT task.
% This is a capacity issue. The model simply cannot fit the relevance data. 
In contrast, GR successfully overcomes this limitation, as evidenced by the near-perfect performance on the original LIMIT dataset.
We believe that this is a crucial evidence for the potential of GR models.

% However, we provide evidence that GR suffers from identifier ambiguity. Unlike the representation collapse in DR, the failure in GR is not due to a lack of memory capacity, but rather a semantic collision in the identifier space. When valid identifiers (like ngrams) are shared across semantically distinct documents, the decoding mechanism cannot distinguish relevance since it lacks context.

% moved from
% 
% by scoring only the most relevant ngrams, 
% ... section 8
% Q4 
% Q1 (COPIED) 
% Q2 (COPIED) 

%

% bullet points
% not too long
% mention that we *know* how to adapt seal to solve limit-h and limit-hs (bias scoring to reward properties of original documents), but that it would not be a general solution to semantic representation.

% \paragraph{Recall-Precision Trade-off}
% On LIMIT (Table~\ref{tab:results_limit}), we observed that the default configuration SEAL/MINDER, which includes all partially decoded identifiers, has a worse R@2, but marginally better R@10 and R@100 compared to standard beam search. This suggests that aggregating more lower-scored ngrams helps broaden the search, but leads to a less clear signal for ranking the most relevant candidates.

\paragraph{GR outperforms BM25}
While BM25 achieved strong recall on the original LIMIT dataset, its performance degraded more rapidly than GR models after the introduction of negatives in LIMIT-H.
These results indicate that BM25's initial success was dependent on the near one-to-one mapping between query terms and relevant documents in the base dataset.
Furthermore, on LIMIT-H the model frequently ranked negatives higher than relevant documents, likely driven by the increased term frequency of common query tokens (e.g., ``Who'', ``likes'') in the negative samples. Interestingly, BM25 showed less sensitivity to scaled ambiguity (LIMIT-HS) than GR models, possibly because the increased document frequency due to negative duplicates counteracted further displacement of relevant documents in the ranking results.
Future GR models should aim to incorporate similar kind of regularization into their scoring approaches.

\paragraph{GR and Identifier Ambiguity}
Our quantitative error analysis (see Tables~\ref{tab:ngram-analysis-LM+FM},~\ref{tab:ngram-analysis-BEAM},~\ref{tab:identifier-analysis-MINDER-NG+PQ}) reveals the fundamental failure mode of GR: in the presence of negatives, the models failed to generate any identifiers unique to the relevant documents.
Consequently, the GR methods could not distinguish between relevant and hard negative documents.
% The system effectively could not perform targeted retrieval and successful retrieval in these settings reduced to a combination of random chance and scoring heuristics that incidentally favor relevant documents.
Our LIMIT-H benchmark provides a critical pivot for the field.
By moving the focus from structural capacity (the vector bottleneck) to fine-grained semantic discrimination, we have identified the ``second wall'' for GR.

\paragraph{GR Scoring Strategy}
% Results on the harder LIMIT variants (Table~\ref{tab:semantic_robustness}) confirm that performance of the evaluated GR methods drops off substantially when the generated identifiers become ambiguous.
% In the default configuration we observe a collapse in R@2.
SEAL default decoding leads to many low-scoring n-grams that inflate the scores for our hard negative documents, which then displace relevant documents to lower ranks. The issue is not document length in itself, but that negative documents tend to contain terms that are semantically related to the item (and consequently the query), but irrelevant to the specific information need. In effect, the cumulative score of identifiers unique to the irrelevant set ($N_i^{I \backslash R}$) outweighs the relevant ngrams.
By restricting scoring to only the highest-probability candidates, the BEAM configuration filters out the noise introduced by these weakly related but irrelevant terms.

% \paragraph{Synthetic Benchmarking and Trivial Solutions}
% The specific construction of our benchmarks implies a trivial (though not generalizable) solution: simply tuning the scoring function to reward properties specific to relevant documents (e.g., shorter document length) would artificially boost R@2. We acknowledge that injecting the full nuance of real-world semantics into a controlled synthetic environment like LIMIT is inherently challenging. Nevertheless, these benchmarks serve as a diagnostic tool, exposing that retrieval with SEAL and MINDER is severely impacted by identifier ambiguity.

\section{Conclusion} \label{sec:conclusion}

In this paper, we systematically evaluated zero-shot generalization capabilities of Generative Retrieval (GR) models, specifically SEAL and MINDER, against the LIMIT benchmark to assess their resilience to the vector bottleneck inherent in embedding-based retrieval.
While our experimental results confirm that GR’s autoregressive architecture successfully bypasses these limits of dense models, we expose an important vulnerability of GR: identifier ambiguity.
% collision
By introducing LIMIT-H and LIMIT-HS—adversarial extensions featuring semantically irrelevant distractors with high lexical overlap — we demonstrate that these GR models struggle with semantic robustness, often failing to distinguish between relevant documents and their hard negative counterparts.

% limitations
We conducted controlled experiments using synthetic data that allowed us to focus on specific failure modes and efficiently collect important insights into the limitations of different IR models. 
In future work, we plan to extend this evaluation to real-world data to confirm our hypothesis.
We also intend to evaluate more advanced models that build upon SEAL, such as LTRGR~\cite{ltrgr}, and more recent models specifically designed for zero-shot settings, such as ZeroGR~\cite{zerogr}.

\balance

\begin{acks}
This work has been funded by the Vienna Science and Technology Fund (WWTF) under the Grant ID 10.47379/VRG24013
\end{acks}

\bibliographystyle{ACM-Reference-Format}
\bibliography{bibliography}

% \newpage
%\appendix

%\input{appendix/configurations}
%\input{appendix/prompt_generate_decoy}

\end{document}